# Analytical Solution to the Fractional Polytropic Gas Spheres

Mohamed I. Nouh[1,2] and Emad A-B. Abdel-Salam[3]

[1]Department of Physics, College of Science, Northern Border University, Arar, Saudi Arabia.

Email: abdo_nouh@hotmail.com

[2]Department of Astronomy, National Research Institute of Astronomy and Geophysics, Helwan, Cairo17211, Egypt.

[3]Department of Mathematics, Faculty of Science, Assiut University, New Valley Branch, El-Kharja 72511, Egypt.

**Abstract.** Lane-Emden equation could be used to model stellar interiors, star clusters and many configurations in astrophysics. Unfortunately, there is an exact solution only for the polytropic index n=0, 1 and 5. In the present paper, a series solution for the fractional Lane-Emden equation is presented. The solution is performed in the frame of modified Rienmann Liouville derivatives. The obtained results recover the well-known series solutions when $\alpha = 1$. Fractional model of n=3 has been calculated and mass-radius relation, density ratio, pressure ratio and temperature ratio have been investigated. We found that the fractional star has a smaller volume and mass than that of the integer star.

1. ## Introduction

Lane–Emden equation (LEE) is a Poisson equation that describes the gravitational potential of the self-gravitating gas sphere. It introduces a simple model to some problems arising in astrophysics such as galactic dynamics and stellar structure.

LEE has exact solutions only for the polytropic indices n=0, 1 and 5, for the rest, the equation is solved by numerical or analytical methods i.e. homotopy perturbation method (Chowdhury and Hashim, 2009), variational iteration method (Ibrahim and Darus, 2008), Sinc-Collocation method (Podlubny, 1999) and an implicit series solution (Momani and Ibrahim, 2008).

The main interest when solving the Lane-Emden equation using power series is how to make the series converges to the outer surface of a gas sphere. Nouh (2004) uses two accelerating techniques to improve the radius of convergence for the range of the polytropic index $0 \leq n < 5$. Hunter (2001) accelerated the series to the surface of the sphere by using Euler transformation.

In the last decade, there is a great effort to use fractional calculus in many areas of physics and astrophysics. Modeling using fractional differential equations gives good results in physics, like anomalous diffusion, signal processing and quantum mechanics (Podlubny 1999; Sokolov et al. 2002; Kilbas et al. 2006; Laskin, 2000; Abdel-Salam and Gumma 2015; Abdel-Salam and Hassan 2015; Abdel-Salam and Hassan 2016; Abdel-Salam et al 2016).

. There are many equations in astrophysics that might be reconstructed and solved in fractional form. El-Nabulsi (2011), who considered a fractional equation of state for white dwarf stars; Bayian and Krisch (2015) have investigated the fractional versions of the stellar structure equations for non-radiating spherical objects. Recently Abdel-Salam and Nouh (2016) introduced a series solution to the fractional isothermal gas sphere and obtained a solution converges to the surface of the sphere with a few series terms when compared with the solution of the integer differential equation.

In this article, we develop an approximate solution for the fractional polytropic gas sphere. A recurrence relation for the coefficient of the power series has been constructed. The structure of the paper is as follows. In section 2, principles of the fractional derivatives are introduced. The fractional Lane-Emden equation is demonstrated in secion 3. The series solution to the fractional polytropic gas sphere is described in section 4. In section 5 we present our results and the conclusion is outlined in section 5.

2. **The Modified Riemann-Liouville Fractional Derivative**

Fractional derivatives have many kinds, of them, are; Riemann–Liouville, Caputo, Kolwankar–Gangal, Oldham and Spanier, Miller and Ross, Cresson's, Grunwald-Letnikov, and modified Riemann–Liouville, Mainardi (2010) and Herrmann (2014).

The modified Riemann–Liouville derivative is written as (Jumarie, 2010)

$$D_x^\alpha f(x) = \begin{cases} \dfrac{1}{\Gamma(-\alpha)} \int_0^x (x-\xi)^{-\alpha-1}[f(\xi)-f(0)]d\xi, & \alpha < 0 \\[2mm] \dfrac{1}{\Gamma(1-\alpha)} \dfrac{d}{dx} \int_0^x (x-\xi)^{-\alpha}[f(\xi)-f(0)]d\xi, & 0 < \alpha < 1 \\[2mm] \dfrac{1}{\Gamma(n-\alpha)} \dfrac{d^n}{dx^n} \int_0^x (x-\xi)^{n-\alpha-1}[f(\xi)-f(0)]d\xi, & n \le \alpha < n+1,\ n \ge 1. \end{cases} \quad (1)$$

Five of the summarized Jumarie (2010) derivative formulas could be written as

$$D_x^\alpha x^\gamma = \frac{\Gamma(\gamma+1)}{\Gamma(\gamma+1-\alpha)} x^{\gamma-\alpha}, \qquad \gamma > 0, \tag{2}$$

$$D_x^\alpha (c f(x)) = c D_x^\alpha f(x), \tag{3}$$

$$D_x^\alpha [f(x)g(x)] = g(x) D_x^\alpha f(x) + f(x) D_x^\alpha g(x), \tag{4}$$

$$D_x^\alpha f[g(x)] = f_g'[g(x)] D_x^\alpha g(x), \tag{5}$$

and

$$D_x^\alpha f[g(x)] = D_g^\alpha f[g(x)](g_x')^\alpha, \tag{6}$$

c is a constant.

Equations (4) and (6) are a direct result from

$$D_x^\alpha f(x) \cong \Gamma(\alpha+1) D_x f(x). \tag{7}$$

He et al. (2012) modified the chain rule (Equation (5)) to

$$D_x^\alpha f[g(x)] = \sigma_x f_g'[g(x)] D_x^\alpha g(x) \tag{8}$$

Modified forms of Equations (4) and (6) could be written as

$$D_x^\alpha [f(x)g(x)] = \sigma_x \{g(x) D_x^\alpha f(x) + f(x) D_x^\alpha g(x)\}, \tag{9}$$

and

$$D_x^\alpha f[g(x)] = \sigma_x D_g^\alpha f[g(x)](g_x')^\alpha, \tag{10}$$

### 3. Fractional Lane-Emden Equation (FLEE)

The polytropic equation of state has the form

$$p = K\rho^\gamma, \tag{11}$$

where
$$\gamma = 1 + \frac{1}{n}.$$

where $n$ is the polytropic index and $K$ is called the pressure constant. The equilibrium structure of a self-gravitating object is derived from the equations of hydrostatic equilibrium. The simplest case is that of a spherical, non-rotating, static configuration, where for a given equation of state all macroscopic properties are parameterized by a single parameter, for example, the central density.

The fractional form of equations of mass conservation and hydrostatic equilibrium are given by

$$\frac{d^\alpha M(r)}{dr^\alpha} = 4\pi r^{2\alpha} \rho, \tag{12}$$

and

$$\frac{d^\alpha P(r)}{dr^\alpha} = -\frac{G M(r)}{r^{2\alpha}} \rho. \tag{13}$$

Rearrange Equation (13) we get

$$\frac{r^{2\alpha}}{\rho} \frac{d^\alpha P(r)}{dr^\alpha} = -G M(r), \tag{14}$$

By performing the first fractional derivative of Equation (14) we get

$$\frac{d^\alpha}{dr^\alpha}\left(\frac{r^{2\alpha}}{\rho} \frac{d^\alpha P(r)}{dr^\alpha}\right) = -G \frac{d^\alpha M(r)}{dr^\alpha}. \tag{15}$$

Combining Equations (11) and (15) we get

$$\frac{d^\alpha}{dr^\alpha}\left(\frac{r^{2\alpha}}{\rho} \frac{d^\alpha P(r)}{dr^\alpha}\right) = -4\pi G r^{2\alpha} \rho, \tag{16}$$

or

$$\frac{1}{r^{2\alpha}} \frac{d^\alpha}{dr^\alpha}\left(\frac{r^{2\alpha}}{\rho} \frac{d^\alpha P(r)}{dr^\alpha}\right) = -4\pi G \rho. \tag{17}$$

Now, by defining the dimensionless function $u$ (Emden function) as

$$\rho = \rho_c u^n, \tag{18}$$

where $\rho$ and $\rho_c$ are the density and central density respectively. The dimensionless variable $x$ could be written as

$$x^\alpha = \frac{r^\alpha}{a}. \tag{19}$$

Inserting Equations (11) and (18) in Equation (17) we get

$$\frac{1}{(ax^\alpha)^2} \frac{d^\alpha}{d(ax^\alpha)} \left( \frac{(ax^\alpha)^2}{\rho_c u^n} \frac{d^\alpha (K\rho^\gamma)}{d(ax^\alpha)} \right) = -4\pi G \rho_c u^n, \tag{20}$$

$$\frac{K}{(ax^\alpha)^2} \frac{d^\alpha}{d(ax^\alpha)} \left( \frac{(ax^\alpha)^2}{\rho_c u^n} \frac{d^\alpha (\rho_c u^n)^{1+\frac{1}{n}}}{d(ax^\alpha)} \right) = -4\pi G \rho_c u^n. \tag{21}$$

The fractional derivative of the Emden function $u$ could be written as

$$\frac{d^\alpha}{dx^\alpha} u^{n+1} = (n+1) u^n \frac{d^\alpha u}{dx^\alpha}. \tag{22}$$

Inserting Equation (22) in Equation (21) we get

$$\frac{K}{a^2 x^{2\alpha}} \frac{d^\alpha}{dx^\alpha} \left( \frac{(n+1) x^{2\alpha} \rho_c^{1+\frac{1}{n}} u^n}{\rho_c u^n} \frac{d^\alpha u}{dx^\alpha} \right) = -4\pi G \rho_c u^n, \tag{23}$$

rearrange

$$\frac{K(n+1)\rho_c^{\frac{1}{n}-1}}{4\pi G a^2} \frac{1}{x^{2\alpha}} \frac{d^\alpha}{dx^\alpha} \left( x^{2\alpha} \frac{d^\alpha u}{dx^\alpha} \right) = -u^n. \tag{24}$$

Now by taking

$$a^2 = \frac{K(n+1)\rho_c^{\frac{1}{n}-1}}{4\pi G}, \tag{25}$$

then the Lane-Emden equation in its fractional form is given by

$$\frac{1}{x^{2\alpha}} \frac{d^\alpha}{dx^\alpha} \left( x^{2\alpha} \frac{d^\alpha u}{dx^\alpha} \right) = -u^n. \tag{26}$$

## 4. Series Solution of the Fractional LE

### 4.1. Successive Fractional Derivatives of the Emden Function

The fractional form of Equation (26) write

$$x^{-2\alpha} D_x^\alpha \left( x^{2\alpha} D_x^\alpha \right) u + u^n = 0, \tag{27}$$

with the initial conditions

$$u(0) = 1, \qquad D_x^\alpha u(0) = 0 \tag{28}$$

where $u = u(x)$, is the Emden function and $0 < \alpha \leq 1$.

By assuming the transform $X = x^\alpha$, the solution could be expressed in a series form as

$$u(X) = \sum_{m=0}^\infty A_m X^m, \tag{29}$$

Using the first initial condition (Equation (28)) to Equation (29) we get $A_0 = 1$ and applying Equation (2) and (4) to Equation (29) we get

$$D_x^\alpha u = A_1 \Gamma(\alpha+1) + \frac{A_1 \Gamma(2\alpha+1) x^\alpha}{\Gamma(\alpha+1)} + \frac{A_1 \Gamma(3\alpha+1) x^{2\alpha}}{\Gamma(2\alpha+1)} + \ldots \tag{30}$$

Applying the second initial condition we obtain

$$D_x^\alpha u(0) = A_1 \Gamma(\alpha+1)$$

Then, $A_1 = 0$ and the series expansion could be written as

$$u(X) = 1 + \sum_{m=2}^\infty A_m X^m \tag{31}$$

Apply the second derivative of the Emden function $u$, we get

$$D_x^{\alpha\alpha} u = D_x^\alpha D_x^\alpha u = A_2 \Gamma(2\alpha+1) + \frac{A_3 \Gamma(3\alpha+1) x^\alpha}{\Gamma(\alpha+1)} + \frac{A_4 \Gamma(4\alpha+1) x^{2\alpha}}{\Gamma(2\alpha+1)} + \ldots \tag{32}$$

At $X = 0$ we have

$$D_x^{\alpha\alpha} u(0) = A_2 \Gamma(2\alpha+1),$$

Applying $j$ times derivatives to the last equation we obtain

$$D^{\overbrace{\alpha \ldots \alpha}^{j \text{ times}}} u(0) = D_x^\alpha \ldots D_x^\alpha u(0) = A_j \Gamma(j\alpha+1),$$

where $A_j$ are constants to be determined.

Now suppose that

$$u^n = G(X) = \sum_{m=0}^{\infty} Q_m X^m, \tag{33}$$

At $X = 0$ we have after $j$ times derivatives

$$D_x^\alpha ... D_x^\alpha G(0) = Q_j \Gamma(j\alpha + 1).$$

## 4.2. Fractional Derivative of Emden Function Raised to Powers

To obtain the fractional derivative of the Emden function $u^n$, we apply the fractional derivative of the product of two functions $u^2$, it will be considered as $u$ times $u$. Similarly, $u^3$ will be considered as $u$ times $u^2$ and so on, Nouh and Saad (2013). Taking the fractional derivative for both sides of Equation (33), we have

$$D_x^\alpha u^n = D_x^\alpha G,$$

that is

$$n u^{n-1} D_x^\alpha u = D_x^\alpha G.$$

or

$$n\, u^n D_x^\alpha u = u\, D_x^\alpha G. \tag{34}$$

Differentiating both sides of Equation (34) $k$ times we have

$$D_x^\alpha ... D_x^\alpha [n G D_x^\alpha u] = D_x^\alpha ... D_x^\alpha (u D_x^\alpha G),$$

Then we have

$$n \sum_{j=0}^{k} \binom{k}{j} u^{\overbrace{\alpha\ ...\alpha}^{j+1\ \text{times}}} G^{\overbrace{\alpha\ ...\alpha}^{k-j\ \text{times}}} = \sum_{j=0}^{k} \binom{k}{j} G^{\overbrace{\alpha\ ...\alpha}^{j+1\ \text{times}}} u^{\overbrace{\alpha\ ...\alpha}^{k-j\ \text{times}}},$$

at $X = 0$, we have

$$n \sum_{j=0}^{k} \binom{k}{j} u^{\overbrace{\alpha\ ...\alpha}^{j+1\ \text{times}}}(0) G^{\overbrace{\alpha\ ...\alpha}^{k-j\ \text{times}}}(0) = \sum_{j=0}^{k} \binom{k}{j} G^{\overbrace{\alpha\ ...\alpha}^{j+1\ \text{times}}}(0) u^{\overbrace{\alpha\ ...\alpha}^{k-j\ \text{times}}}(0),$$

or

$$n \sum_{j=0}^{k} \binom{k}{j} A_{j+1} \Gamma((j+1)\alpha + 1) Q_{k-j} \Gamma(\alpha(k-j)+1) = \sum_{j=0}^{k} \binom{k}{j} Q_{j+1} \Gamma((j+1)\alpha + 1) A_{k-j} \Gamma(\alpha(k-j)+1),$$

So, we get the following equations

$$n\sum_{j=0}^{k} \frac{k!\Gamma(\alpha(k-j)+1)\Gamma((j+1)\alpha+1)}{j!(k-j)!} A_{j+1}Q_{k-j} = \sum_{j=0}^{k} \frac{k!\Gamma((j+1)\alpha+1)\Gamma(\alpha(k-j)+1)}{j!(k-j)!} A_{k-j}Q_{j+1},$$

$$n\sum_{j=0}^{k} \frac{k!\Gamma(\alpha(k-j)+1)\Gamma((j+1)\alpha+1)}{j!(k-j)!} A_{j+1}Q_{k-j}$$
$$= \sum_{j=0}^{k-1} \frac{k!\Gamma((j+1)\alpha+1)\Gamma(\alpha(k-j)+1)}{j!(k-j)!} A_{k-j}Q_{j+1} + \Gamma((k+1)\alpha+1)A_0 Q_{k+1},$$

and

$$\Gamma((k+1)\alpha+1)A_0 Q_{k+1} = n\sum_{j=0}^{k} \frac{k!\Gamma(\alpha(k-j)+1)\Gamma((j+1)\alpha+1)}{j!(k-j)!} A_{j+1}Q_{k-j}$$
$$- \sum_{j=0}^{k-1} \frac{k!\Gamma((j+1)\alpha+1)\Gamma(\alpha(k-j)+1)}{j!(k-j)!} A_{k-j}Q_{j+1}.$$

In the last equation, let $i = j + 1$ in the first summation and $i = k - j$ in the second summation, we get

$$\Gamma((k+1)\alpha+1)A_0 Q_{k+1} = n\sum_{i=1}^{k+1} \frac{k!\Gamma(\alpha(k+1-i)+1)\Gamma(i\alpha+1)}{(i-1)!(k+1-i)!} A_i Q_{k+1-i}$$
$$- \sum_{i=1}^{k} \frac{k!\Gamma((k+1-i)\alpha+1)\Gamma(i\alpha+1)}{(k-i)!i!} A_i Q_{k+1-i}$$

If $m = k+1$, then

$$\Gamma(m\alpha+1)A_0 Q_m = n\sum_{i=1}^{m} \frac{(m-1)!\Gamma(\alpha(m-i)+1)\Gamma(i\alpha+1)}{(i-1)!(m-i)!} A_i Q_{m-i}$$
$$- \sum_{i=1}^{m-1} \frac{(m-1)!\Gamma((m-i)\alpha+1)\Gamma(i\alpha+1)}{(m-1-i)!i!} A_i Q_{m-i}$$

or

$$\Gamma(m\alpha+1)A_0 Q_m = n\sum_{i=1}^{m} \frac{(m-1)!\Gamma(\alpha(m-i)+1)\Gamma(i\alpha+1)}{(i-1)!(m-i)!} A_i Q_{m-i}$$
$$- \sum_{i=1}^{m-1} \frac{(m-1)!(m-i)\alpha\Gamma((m-i)\alpha)\Gamma(i\alpha+1)}{(m-1-i)!i!} A_i Q_{m-i}$$

By adding the zero value $\left\{ -\dfrac{(m-m)\alpha\, \Gamma(m\alpha+1)}{m} A_m Q_0 \right\}$ to the second summation, we get

$$\Gamma(m\alpha+1)A_0Q_m = n\sum_{i=1}^{m} \frac{(m-1)!\Gamma(\alpha(m-i)+1)\Gamma(i\alpha+1)}{(i-1)!(m-i)!} A_i Q_{m-i}$$

$$-\sum_{i=1}^{m} \frac{(m-1)!\Gamma((m-i)\alpha+1)\Gamma(i\alpha+1)}{(m-1-i)!i!} A_i Q_{m-i}$$

Then the coefficients $Q_m$ could be written as

$$Q_m = \frac{1}{\Gamma(m\alpha+1)A_0} \sum_{i=1}^{m} \frac{(m-1)!\Gamma(\alpha(m-i)+1)\Gamma(i\alpha+1)}{i!(m-i)!}[in-m+i]A_i Q_{m-i}, \quad \forall\ m \geq 1, \tag{35}$$

where

$$A_0 = 1,\ A_1 = 0,\quad Q_0 = A_0^n = 1,\ Q_1 = \frac{\Gamma(\alpha+1)n}{\Gamma(\alpha+1)A_0} A_1 Q_0 = 0.$$

### 4.3. Fractional Derivatives of the Series Expansion of the Emden Function

Using $u = 1 + \sum_{m=2}^{\infty} A_m X^m$ we obtain

$$D_x^\alpha u = \sum_{m=2}^{\infty} A_m \sigma_{1x} mX^{m-1} \frac{\Gamma(\alpha+1)}{\Gamma(\alpha+1-\alpha)} x^{\alpha-\alpha}$$

$$= \sum_{m=2}^{\infty} A_m mX^{m-1} \frac{\Gamma(\alpha+1)}{\Gamma(\alpha+1-\alpha)} \frac{\Gamma(m\alpha+1)}{m\Gamma(\alpha+1)\Gamma(m\alpha+1-\alpha)} \tag{36}$$

$$= \sum_{m=2}^{\infty} A_m X^{m-1} \frac{\Gamma(m\alpha+1)}{\Gamma(m\alpha+1-\alpha)},$$

where $\sigma_{1x} = \frac{\Gamma(m\alpha+1)}{m\Gamma(\alpha+1)\Gamma(m\alpha+1-\alpha)}$, see Abdel-Salam and Nouh (2016).

The second derivative of the Emden function $u$ could be given by

$$D_x^\alpha D_x^\alpha u = \sum_{m=2}^{\infty} A_m \sigma_{2x}(m-1)X^{m-2} \frac{\Gamma(m\alpha+1)}{\Gamma(m\alpha+1-\alpha)} \frac{\Gamma(\alpha+1)}{\Gamma(\alpha+1-\alpha)} x^{\alpha-\alpha} = \sum_{m=2}^{\infty} X^{m-2} \frac{A_m \Gamma(m\alpha+1)}{\Gamma(m\alpha+1-2\alpha)}, \tag{37}$$

where $\sigma_{2x} = \frac{\Gamma((m-1)\alpha+1)}{(m-1)\Gamma(\alpha+1)\Gamma((m-1)\alpha+1-\alpha)}$, see Abdel-Salam and Nouh (2016).

Substituting Equations (36) and (37) in Equation (29) we have

$$x^{2\alpha}\sum_{m=2}^{\infty} X^{m-2} \frac{A_m \Gamma(m\alpha+1)}{\Gamma(m\alpha+1-2\alpha)} + \frac{\Gamma(2\alpha+1)x^\alpha}{\Gamma(\alpha+1)} \sum_{m=2}^{\infty} X^{m-1} \frac{A_m \Gamma(m\alpha+1)}{\Gamma(m\alpha+1-\alpha)} + x^{2\alpha}\left[1 + \sum_{m=2}^{\infty} Q_m X^m\right] = 0,\ ,$$

or

$$\sum_{m=2}^{\infty} A_m X^m \left[ \frac{\Gamma(m\alpha+1)}{\Gamma(m\alpha+1-2\alpha)} + \frac{\Gamma(2\alpha+1)\Gamma(m\alpha+1)}{\Gamma(\alpha+1)\Gamma(m\alpha+1-\alpha)} \right] + X^2 + \sum_{m=2}^{\infty} Q_m X^{m+2} = 0. \tag{38}$$

By putting $m = k+2$ in the first part and $m = k$ in the third part of Equation (38), we get

$$\sum_{k=0}^{\infty} A_{k+2} X^{k+2} \left[ \frac{\Gamma((k+2)\alpha+1)}{\Gamma(k\alpha+1)} + \frac{\Gamma(2\alpha+1)\Gamma((k+2)\alpha+1)}{\Gamma(\alpha+1)\Gamma(k\alpha+1+\alpha)} \right] + X^2 + \sum_{k=2}^{\infty} Q_k X^{k+2} = 0,$$

After some manipulations, we get the recurrence relation of the coefficients as

$$\left[ \frac{\Gamma((k+2)\alpha+1)}{\Gamma(k\alpha+1)} + \frac{\Gamma(2\alpha+1)\Gamma((k+2)\alpha+1)}{\Gamma(\alpha+1)\Gamma(k\alpha+1+\alpha)} \right] A_{k+2} + Q_k = 0 \quad \forall \, k \geq 2 ,$$

### 4.4. The Recurrence Relations

Now we can calculate the coefficients of the series expansion from the following two recurrence relations

$$A_{k+2} = -Q_k \frac{\Gamma(\alpha+1)\Gamma(k\alpha+1)\Gamma(k\alpha+\alpha+1)}{\Gamma(k\alpha+2\alpha+1)[\Gamma(\alpha+1)\Gamma(k\alpha+\alpha+1) + \Gamma(2\alpha+1)\Gamma(k\alpha+1)]}, \quad \forall \, k \geq 2 \tag{39}$$

and

$$Q_m = \frac{\Gamma(m)}{\Gamma(m\alpha+1)A_0} \sum_{i=1}^{m} \frac{\Gamma(\alpha(m-i)+1)\Gamma(i\alpha+1)}{\Gamma(i+1)\Gamma(m-i+1)} [in - m + i] A_i Q_{m-i}, \quad \forall \, m \geq 1 \tag{40}$$

If we put $\alpha = 1$ in Equations (39) and (40) the series coefficients are reduced to the well-known series solution of the integer LEE. For $k = 0, 1, 2, 3$ in Equations (39) and (40) we get

$$A_2 = -\frac{[\Gamma(\alpha+1)]^2}{\Gamma(2\alpha+1)\left[[\Gamma(\alpha+1)]^2 + \Gamma(2\alpha+1)\right]},$$

$$A_3 = 0,$$

$$A_4 = \frac{n\Gamma(\alpha+1)^3 \Gamma(2\alpha+1)\Gamma(3\alpha+1)}{\Gamma(4\alpha+1)\Gamma(2\alpha+1)\left[\Gamma(\alpha+1)\Gamma(3\alpha+1) + \Gamma(2\alpha+1)^2\right]\left[\Gamma(\alpha+1)^2 + \Gamma(2\alpha+1)\right]},$$

$$A_5 = 0.$$

Then the series solution at $\alpha = 1$ is reduced to the well-known solution of Equation (27) as

$$u_n(x) = 1 - \frac{1}{6} x + \frac{n}{120} x^2 - \ldots\ldots\ldots\ldots$$

## 5. Results

A MATHEMATICA program has been constructed to calculate the series coefficients for the range of the polytropic index $0 \leq n < 5$. The code has two parts, the first is to determine the radius of convergence of the series and the second is devoted to calculating the polytropic physical parameters i.e. mass, radius, temperature, and density.

To test the code, we first run it for $\alpha = 1$, as it is well known, the series expansion of Lane-Emden diverges for the polytropic index n>1.9. To accelerate the series to the surface of the polytrope, we use the two acceleration techniques of Nouh (2004). Then we calculate the radius of the gas sphere for different values of $\alpha$ for each polytropic index.

In Figs. 1 and 2 we display the Emden function at different values of the fractional parameter $\alpha$ for the polytropic index n=0 and 3 respectively. As we see, the effects of changing the fractional parameter on the first zero ($x1$) are very small for n=0 (a gas sphere with constant density). For n=3, a small change in the fractional parameter leads to a remarkable change in the first zero.

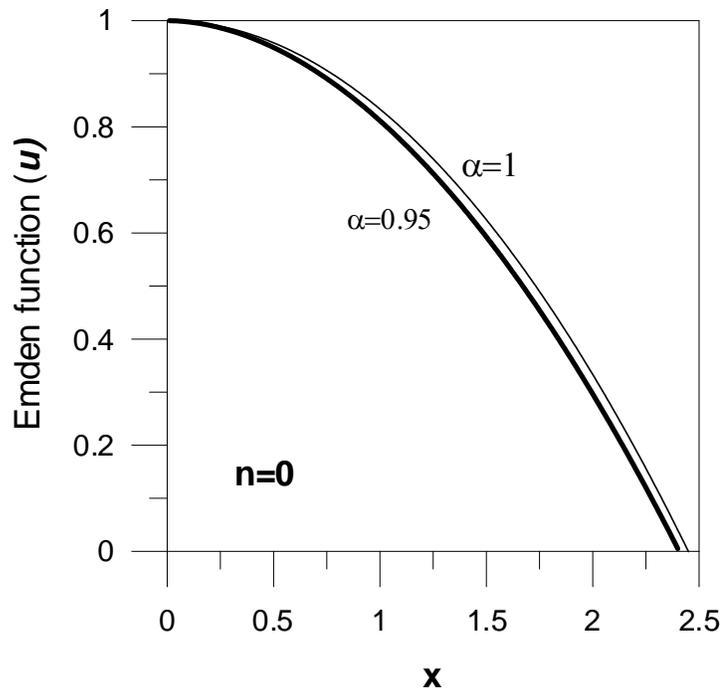

**Figure 1. Emden function for the fractional polytrope with n=0.**

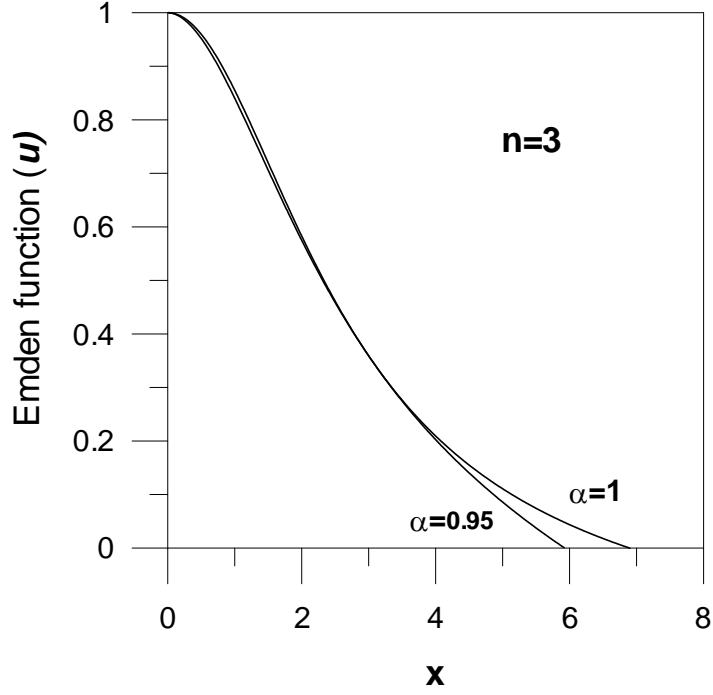

**Figure 2. Emden function of the fractional polytrope with n=3.**

The mass contained in a radius $r$ is given by

$$M(r^\alpha) = \int_0^r 4\pi r^{2\alpha} \rho \, dr^\alpha \ . \tag{41}$$

Inserting Equations (18) and (19) for $\rho$ and $r^\alpha$ we found

$$M(x^\alpha) = 4\pi a^3 \rho_c \int_0^x x^{2\alpha} u^n \, dx^\alpha \ , \tag{42}$$

by substituting Equation (26) for the Emden function $u^n$ we get

$$M(x^\alpha) = 4\pi a^3 \rho_c \int_0^x x^{2\alpha} \left[ -\frac{1}{x^{2\alpha}} \frac{d^\alpha}{dx^\alpha} \left( x^{2\alpha} \frac{d^\alpha u}{dx^\alpha} \right) \right] dx^\alpha$$

$$= 4\pi a^3 \rho_c \int_0^x \left[ -\frac{d^\alpha}{dx^\alpha} \left( x^{2\alpha} \frac{d^\alpha u}{dx^\alpha} \right) \right] dx^\alpha$$

$$= 4\pi a^3 \rho_c \left[ -\left( x^{2\alpha} \frac{d^\alpha u}{dx^\alpha} \right) \right]. \tag{43}$$

with $a$ could be followed from Equation (25), then the mass is given by

$$M(x^\alpha) = 4\pi \left[\frac{K(n+1)}{4\pi G}\right]^{\frac{3}{2}} \rho_c^{\frac{3-n}{2n}} \left[-\left(x^{2\alpha} \frac{d^\alpha u}{d x^\alpha}\right)\right]_{x=x_1}. \tag{44}$$

The radius of the polytrope is given by

$$R^\alpha = a\, x_1^\alpha,$$

where $x_1^\alpha$ is the first zero of Lane-Emden function. Inserting $a$ in Equation (45) we get

$$R^\alpha = \left[\frac{K(n+1)}{4\pi G}\right]^{\frac{1}{2}} \rho_c^{\frac{1-n}{2n}} x_1^\alpha. \tag{45}$$

The pressure and temperature of the polytrope are given by

$$P = P_c\, u^{n+1}, \tag{46}$$

and

$$T = T_c\, u^n. \tag{47}$$

To start the calculation of the polytropic physical parameters, we will assume the mass, radius and central temperature of the sun as , $R_0 = 6.9598 \times 10^8\ m$ and $T_c = 1.570 \times 10^7\ K$ respectively. The central density is computed from the equation

$$\rho_c = -\frac{x^{2\alpha} M_0}{4\pi R_0^3 \left[\left(\dfrac{d^\alpha u}{d x^\alpha}\right)\right]_{x=x1}},$$

the results are listed in Table 1, where column 1 is the fractional parameter $\alpha$, column 2 is the first zero of the Emden function and columns 3 and 4 are respectively the ratios of the total radius and total mass of the star to the radius and mass of the sun. As was expected by Bayian and Krisch (2015), the fractional polytropic gas sphere has a volume and mass smaller than the integer one. Figs 3 to 7 display the distributions of the density ratio, mass-radius relation, pressure distribution and temperature distribution for the gas sphere with polytropic index n=3. The results indicate that; the mass distribution of the fractional star will appear much different than that of the integer star, i.e. the fractional object is more compacted than the integer one.

**Table (1): Radius of convergence and physical parameters of the fractional polytrope with n=3.**

| $\alpha$ | x1 | $R_*/R_0$ | $M_*/M_0$ |
|---|---|---|---|
| 1 | 6.9 | 1 | 1 |
| 0.99 | 6.55 | 0.981 | 0.963 |
| 0.98 | 6.32 | 0.963 | 0.928 |
| 0.97 | 6.15 | 0.946 | 0.896 |
| 0.96 | 6.02 | 0.930 | 0.866 |
| 0.95 | 5.92 | 0.918 | 0.842 |

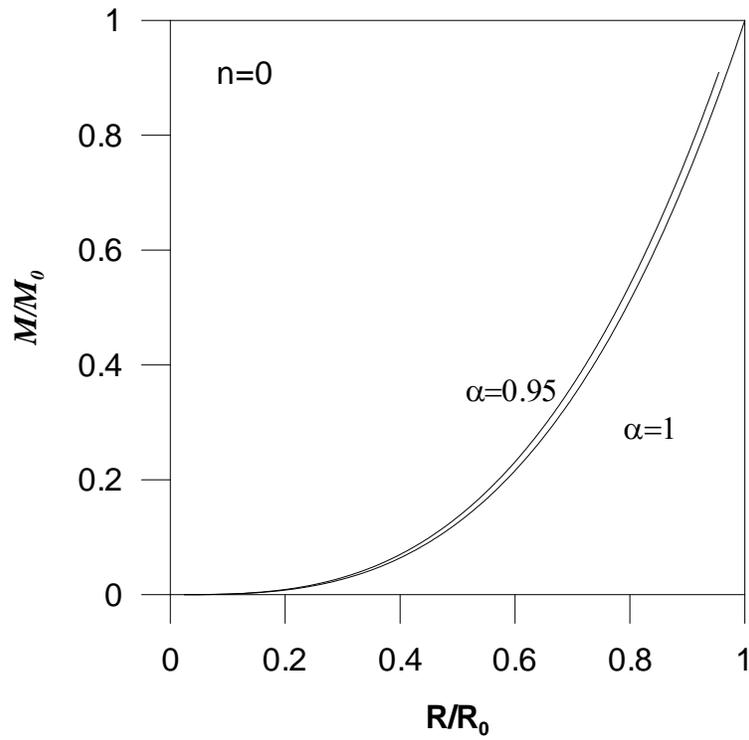

**Fig 3: Mass-Radius relation of the fractional polytrope with n=0.**

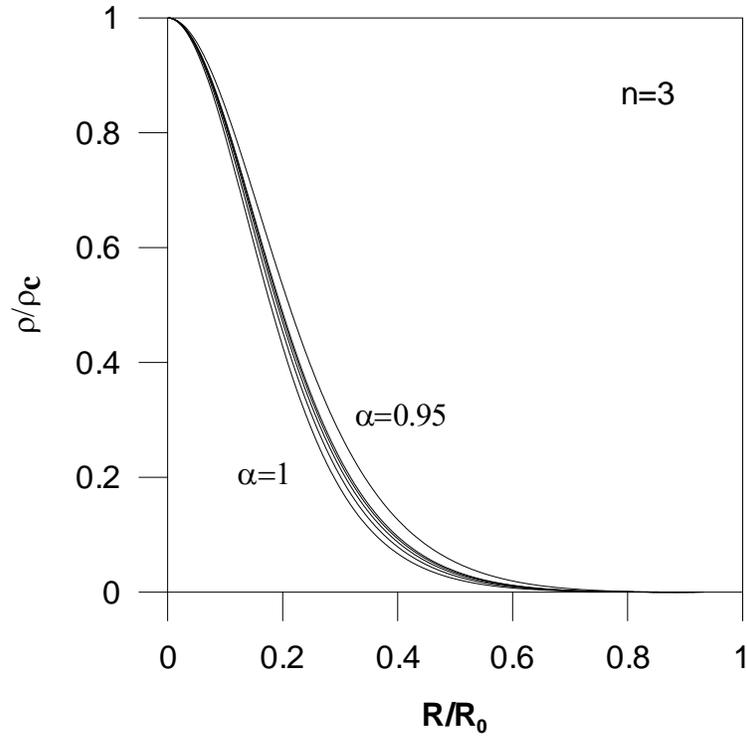

**Fig 4: Distribution of the density ratio of the fractional polytrope with n=3.**

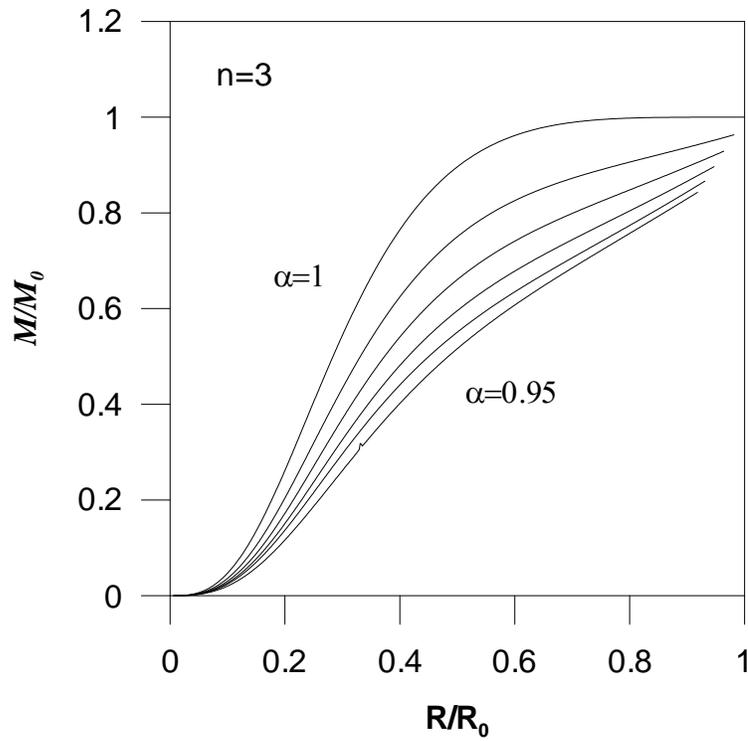

**Fig 5: Mass-Radius relation of the fractional polytrope with n=3.**

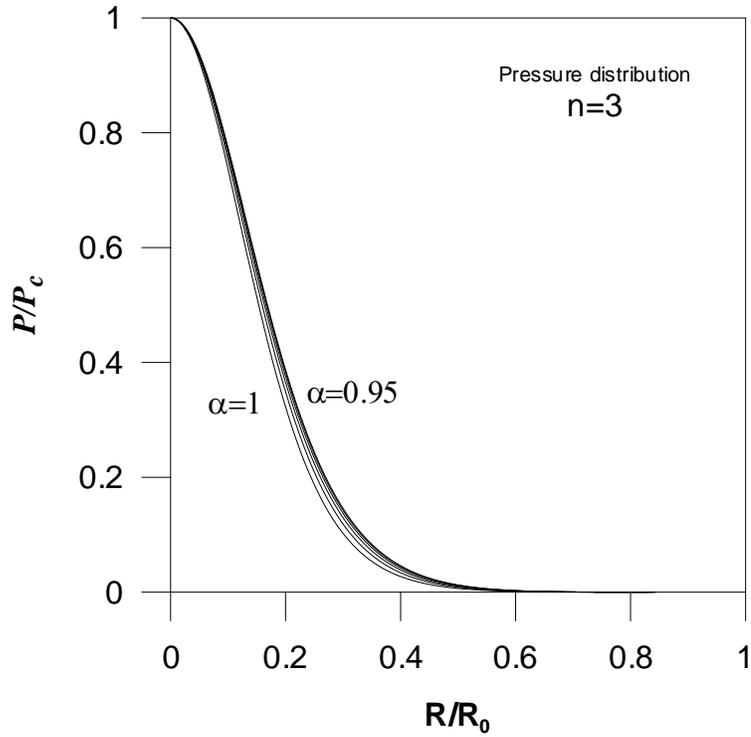

**Fig 6: Pressure distribution of the fractional polytrope with n=3.**

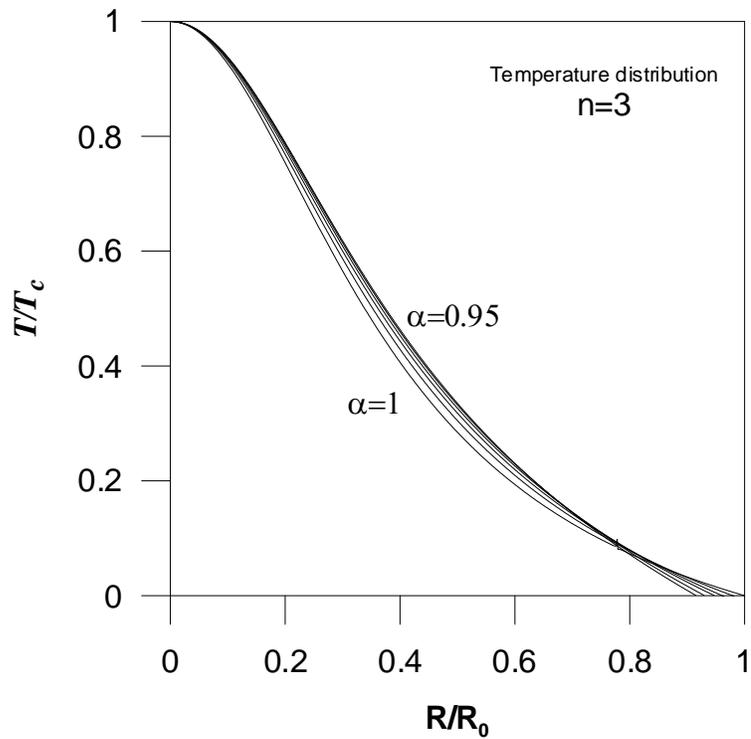

**Fig7: Temperature distribution of the fractional polytrope with n=3.**

## 6. Concluding Remarks

In the present paper, we presented an approximate solution to the fractional Lane-Emden equation. With the help of fractal index, we constructed a recurrence relation for the coefficients of the series expansion of Emden function. The results obtained show that the solution is converged everywhere for the range polytropic index $0 \leq n \leq 4.99$. The dependence of the computed first zero of the Emden function on the fractional parameter $\alpha$ is clear. For $\alpha = 1$, the solution is reduced to the well-known series solution. The results reached in the present article for the fractional polytrope may be considered valuable.

Here, we present application to solar type stars or what called standard stellar model with polytropic index n=3. The effects of the fractional parameter on the physical characteristics of the polytrope like mass radius relation, density profile and pressure profiles is clear. Actually each stellar configuration (with different polytopic indices) will have upper limit of the fractional parameter. For example, compact objects (white dwarfs and neutron stars) with n=1.5 should have smaller mass and volume, therefore the matter inside these objects will be more compacted (compactness ratio). This issue requires intensive calculations which will be treated in a separate paper.